\documentclass[prd,nofootinbib,twocolumn,superscriptaddress,preprintnumbers,balancelastpage,nofootinbib]{revtex4}

\usepackage{amsmath,hyperref,amssymb}
\usepackage{graphicx}
\usepackage{dcolumn}
\usepackage{bm}
\usepackage{verbatim}
\usepackage{tabularx}
\usepackage{slashed}
\usepackage{cancel}
\usepackage{float}

\def\be{\begin{equation}}
\def\ee{\end{equation}}
\def\bea{\begin{eqnarray}}
\def\eea{\end{eqnarray}}

\newcommand{\Eref}[1]{Eq.~\ref{#1}}

\newcommand{\Mev}{{\rm MeV}}

\begin{document}

\title{Gamma-rays flashes from dark photons in neutron star mergers}

\author{Melissa D. Diamond}
\email{mdiamon8@jhu.edu }
\affiliation{Department  of  Physics  and  Astronomy,  Johns  Hopkins  University,  Baltimore,  MD  21218, U.S.A.}

\author{Gustavo Marques-Tavares}
\email{gusmt@umd.edu}
\affiliation{Maryland Center for Fundamental Physics, Department of Physics, University of Maryland, College Park, MD 20742, U.S.A.}

\vspace*{1cm}

\begin{abstract} 

In this letter we begin the study of visible dark sector signals coming from binary neutron star mergers. We focus on dark photons emitted in the 10 ms - 1 s after the merger, and show how they can lead to bright transient gamma-ray signals.  The signal will be approximately isotropic, and for much of the interesting parameter space will be close to thermal, with an apparent temperature of  $\sim100$ keV. These features can distinguish the dark photon signal from the expected short gamma-ray bursts produced in neutron star mergers, which are beamed in a small angle and non-thermal.  We calculate the expected signal strength and show that for dark photon masses in the $1-100$ MeV range it can easily lead to total luminosities larger than $10^{46}$ ergs for much of the unconstrained parameter space. This signal can be used to probe a large fraction of the unconstrained parameter space motivated by freeze-in dark matter scenarios with interactions mediated by a dark photon in that mass range.  We also comment on future improvements when proposed telescopes and mid-band gravitational detectors become operational.
\end{abstract}

\maketitle

\section{Introduction}
The detection of gravitational waves (GW) from binary neutron star (BNS) mergers, and the observation of their electromagnetic counterparts has inaugurated a new era in multi-messenger astronomy~\cite{TheLIGOScientific:2017qsa,GBM:2017lvd,Monitor:2017mdv}. Upcoming observations will shed light on the physics of neutron stars, the origin of heavy elements and models of stellar evolution \cite{Cowan:2019pkx,Meszaros:2001vi,Pian:2020vul,Meszaros:2019xej}. This new window into the Universe offers great potential as a probe of physics beyond the standard model (BSM), and in particular of scenarios involving very weakly interacting new states. 

When two neutron stars merge, a meta-stable remnant with nuclear densities and temperatures in the 10s of MeV forms, similar to the proto-neutron stars formed in core-collapse supernovae \cite{Camelio:2020mdi, Endrizzi:2019trv, Radice:2020ids,Perego:2019adq}. These hot remnants are a promising source of new weakly coupled particles. It has already been shown that such events can produce a large flux of neutrinos \cite{Radice:2018pdn}. Nonetheless, the low merger rate implies that if we are interested in seeing at least $\mathcal{O}(1)$ merger/year, their typical distance will be $\sim 100$ Mpc, making direct observation of the new particles challenging due to their small flux at large distances and small couplings. However, if the new particles are unstable and can decay to visible particles, such as leptons or photons, they can produce very bright signals making BNS mergers a powerful probe of new physics.

There have been a number of proposals to search for new physics affecting the GW signal from binary black-holes or neutron star mergers~\cite{Endlich:2017tqa,Hook:2017psm,Sagunski:2017nzb,Croon:2017zcu,Huang:2018pbu,Bezares:2019jcb,Dror:2019uea,Sennett:2019bpc}, in addition to potential cooling constraints due to emission of weakly coupled BSM particles from the remnant~\cite{Dietrich:2019shr,Harris:2020qim}. In this letter, we initiate the study of dark sector electromagnetic signals in BNS mergers. For concreteness, we focus exclusively on dark photons, though this framework is applicable to other dark sector models. The dark photon is a new massive vector field that kinetically mixes with the photon, and through this mixing interacts with charged standard model matter \cite{Holdom:1985ag,Fabbrichesi:2020wbt}. It corresponds to one of the three renormalizable portals between the Standard Model (SM) and dark sectors, which are sectors not charged under the SM gauge group~\cite{Essig:2013lka,Alexander:2016aln}. Dark matter might be part of such a dark sector,  and its interactions with the standard model mediated by the dark photon may account for the observed dark matter density. The relic abundance can be obtained either through standard freeze-out, corresponding to scenarios in which dark matter reaches thermal equilibrium with the visible sector and requires larger couplings, or through freeze-in in which the interactions are so weak that dark matter never reaches thermal equilibrium with the SM~\cite{Hall:2009bx,Chu:2011be}.

The cross-sections suggested by the freeze-out scenario motivated a large experimental program intended to cover the parameter space corresponding to the observed relic density. In contrast, the freeze-in scenario points to tiny couplings, with the kinetic mixing parameter as small as $10^{-11}$, making it more difficult to probe experimentally. There are a number of recent direct detection proposals aimed at probing the freeze-in parameters for very light dark photons, which take advantage of the cross-section enhancement at low velocities characteristic of interactions mediated by light particles~(e.g. \cite{Essig:2015cda,Knapen:2017ekk,Abramoff:2019dfb,Aralis:2019nfa}). Probing the freeze-in scenario for larger dark photon masses, when there is no significant enhancement of the cross-section, is much more challenging. Most constraints on this scenario come from searches sensitive to the dark photon directly, generally using cosmological or astrophysical signals \cite{Redondo:2008ec,An:2013yfc,Fradette:2014sza,Kazanas:2014mca,Chang:2016ntp,Hardy:2016kme,DeRocco:2019njg,Hong:2020bxo}. We will show that dark photons in the mass range $\sim 1 - 100$ MeV can be produced copiously in the remnant of a neutron star merger. Their decays lead to a transient bright gamma-ray signal that can be used to search for dark photons in much of the remaining viable parameter space for freeze-in with a mediator in that mass range.

\section{Dark Photon production and decay}
\label{sec:production}

We will concentrate on dark photon production and assume that any other new particles are heavy and irrelevant for BNS mergers. The relevant terms in the lagrangian are
\begin{equation}
    \mathcal{L}\supset \frac{1}{2}m'A_{\mu}'A'^{\mu}-\frac{1}{4}F_{\mu \nu}'F'^{\mu\nu}-\frac{\epsilon}{2}F_{\mu\nu}'
    F^{\mu\nu} \, ,
\end{equation}
where $A'^\mu$ is the dark photon, $m'$ its mass, $F^{\mu\nu}$ the photon field strength, and $\epsilon$ the kinetic mixing.

The production of dark photons in the proto-neutron star is dominated by nucleon-nucleon bremsstrahlung, as in the supernova case. The flux of dark photons can be calculated following Ref.~\cite{Chang:2016ntp},
\begin{equation}
\label{eq:number}
 \frac{dN}{dVdt} = \int \frac{d\omega \omega^2 v}{2\pi^2}e^{-\omega/T}(\Gamma_{\text{abs},L} +2\Gamma_{\text{abs},T}),
\end{equation}
where $\omega$ is the frequency of the dark photon, $v$ its velocity, $T$ the local temperature and $\Gamma_{\text{abs}, T/L}$ the absorption width for the transverse/longitudinal dark photon modes. Integrating this over the production volume gives the total number of dark photons produced.  The absorption width can be computed using the soft-radiation approximation, as discussed in Ref.~\cite{Rrapaj:2015wgs}. Ignoring Pauli blocking (which is at most an $\mathcal{O}(1)$ effect), we find
\begin{equation}
\begin{aligned}
    \Gamma_{\text{abs, } X} = & \, \frac{32 \alpha n_{\text{n}} n_{\text{p}} \epsilon^2 \, m^{\prime \, 4}}{3 \pi \omega^3 \left((m^{\prime \, 2} - \text{Re}\Pi_{X})^2 + \text{Im} \Pi_{X}^2\right)} \left( \frac{\pi T}{m_{\text{N}}} \right)^{3/2} \\
    \times & \langle \sigma_\text{np}^{(2)} (T) \rangle \, \times \left\{
    \begin{array}{ll} 
    1 \, , & X = T \\
    (m^\prime/\omega)^2 \, , & X = L
    \end{array}\right. \, ,
\end{aligned}
\end{equation}
where $X = T(L)$ refers to the transverse (longitudinal) polarization of the dark photon, $m_N$ is the nucleon mass, $n_{n/p}$ is the neutron/proton density, $\Pi_X$ is the in medium polarization tensor of the photon (see Supplemental Materials~\cite{supplement} for their explicit form) and $\langle \sigma_\text{np}^{(2)} \rangle$ is the weighted proton neutron scattering cross-setion taken from Ref.~\cite{Rrapaj:2015wgs}. From \Eref{eq:number}, we calculate the dark photon luminosity
\begin{equation}
\label{energy}
  \frac{dE}{dVdt} = \int \frac{d\omega \omega^3 v}{2\pi^2}e^{-\omega/T}(\Gamma_{\text{abs},L} +2\Gamma_{\text{abs},T}) \, .
\end{equation}

\begin{figure}[th]
    \centering
    \includegraphics[width=0.46\textwidth]{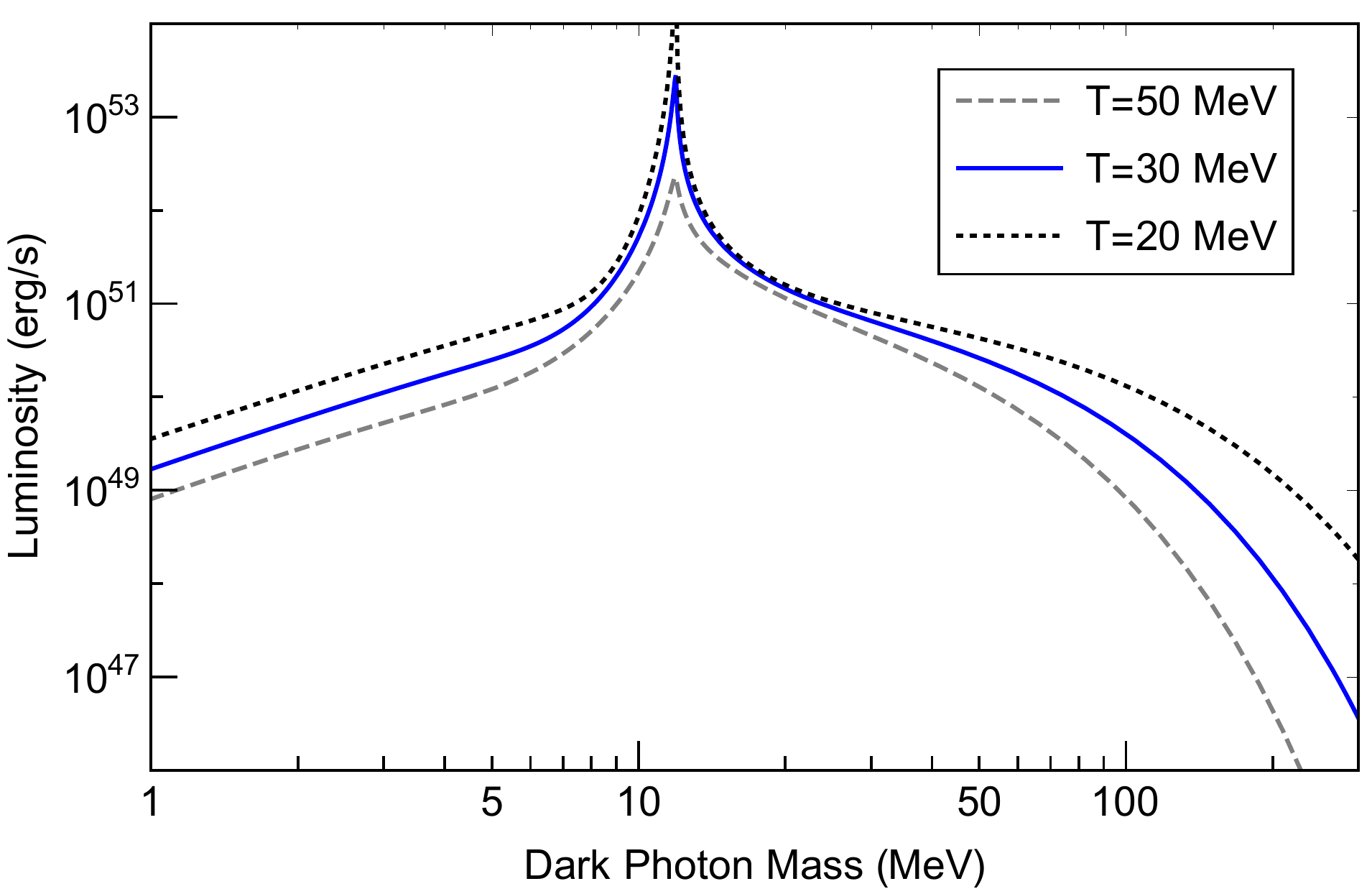}
    \caption{Luminosity of dark photons with $\epsilon = 10^{-10}$,  produced in BNS mergers using the simplified profile. We use $T=30$ MeV (blue curve) for calculations in this work.  The dashed and dotted lines show luminosity for other values of temperature. The enhancement near $10$ MeV comes from resonant mixing between the dark photon and the photon.}
    \label{fig:dar_photon_lum}
\end{figure}

Based on BNS merger simulations in Ref.~\cite{Perego:2019adq,Radice:2020ids,Bernuzzi:2020txg,Endrizzi:2019trv}, we assume a simplified, spherically symmetric description of the merger remnant, with a constant temperature, density and electron fraction: $T=30$ MeV, $\rho = 4\times 10^{14}$ g cm$^{-3}$,  $Y_e = 0.1$ in the region 5 km-10 km from the center of the remnant. We consider only the dark photons produced in this hot region and ignore contributions from other colder regions. In Fig.~\ref{fig:dar_photon_lum}, we show the luminosity of dark photons as a function of mass and  how it varies with the remnant's temperature.

The time evolution of the remnant depends on the mass of the original neutron stars and the equation of state of nuclear matter. Different initial conditions can produce a variety of different remnants that persist for between $1-1000$ ms before collapsing to a black hole \cite{Perego:2019adq,Camelio:2020mdi,Endrizzi:2019trv,Bernuzzi:2020txg,Radice:2020ids}. Light remnants might not collapse at all, instead forming a heavier neutron star \cite{Fujibayashi:2020dvr}. Nonetheless, cooling will decrease their temperature from $T\sim 30$ MeV within a few seconds~\cite{Dietrich:2019shr,Radice:2020ddv}.  To illustrate this range of possibilities, we  present results assuming two different scenarios for dark photon emissions: emissions lasting $10$ ms after the merger and emissions lasting $1$ s after the merger. Note that most analyses of GW170817 conclude that the remnant must have lasted for at least $10$ ms \cite{Murguia-Berthier:2020tfs,Shibata:2019ctb,Shibata:2017xdx,Ruiz:2017due}. For simplicity, we ignore  gravitational redshift, which depends strongly on the density distribution in the central region of the merger~(we estimate its impact in the Supplemental Material~\cite{supplement}).

After production, the dark photons decay to electron-positron pairs, forming an expanding plasma shell. The dark photon decay width at rest is
\begin{equation}
    \Gamma = \frac{1}{3}\alpha\epsilon^2 m'\sqrt{1-\frac{4m_{\text{e}}^2}{m'^2}}\left( 1+\frac{2m_{\text{e}}^2}{m'^2}\right) \, ,
\end{equation}
where $m_e$ is the electron mass. The initial Lorentz factor of the expanding shell is approximately the average Lorentz factor of the dark photon flux,
\begin{equation}
    \gamma_0=\frac{\langle \omega \rangle}{m'} \, ,
    \label{eq: initial boost}
\end{equation}
where $\langle \omega \rangle$ is the average dark photon energy.  In the merger frame, the dark photon decay length is 
\begin{equation}
    d = \frac{ \gamma_0 v}{\Gamma} \, .
    \label{eq: decay length}
\end{equation}
The width of the plasma shell immediately after the decay is
\begin{equation}
\label{width}
    \delta =\frac{1}{\gamma_0 \Gamma} \, , \qquad \delta'=\frac{1}{\Gamma} 
\end{equation}
in the star frame and plasma frame respectively.

The photon signal  depends on the evolution of this plasma. For simplicity, we focus on parameters such that the dark photons decay at least 1000 km from the center of the merger, where ambient baryon density and magnetic fields can be safely neglected.~\footnote{
Magnetic fields do not directly affect the dark photons, but would lead to non-trivial dynamics for the electron-positron plasma formed after the decay.
}  When calculating the number of leptons that results from dark photon decays, we only include the fraction coming from decays at least 1000 km away from the merger. Dark photons in most of the viable parameter space have long enough decay lengths that this requirement does not cause a significant change in sensitivity.

\section{Gamma-ray signal}
\label{sec:signals}

We are interested in the photon signal arising from the the electrons and positrons produced by dark photon decays. With our simplified approximation for the remnant, the signal will be isotropic\footnote{Realistically the remnant will not be spherically symmetric and will have $\mathcal{O}(1)$ variations in the luminosity depending on the inclination angle with respect to the plane of the merger.}, 
peaked between 100 keV and 10s of MeV and visible within about a second of the merger. The isotropic nature of the signal, its duration and its spectral information can be used to distinguish it from the short gamma-ray burst (GRB) expected to be produced from relativistic jets after the remnant collapses to a black hole \cite{Monitor:2017mdv,Meszaros:2006rc}.

We track the dynamics of the plasma starting from a radius equal to the decay length of the dark photon in the star frame, given by Eq.~\ref{eq: decay length}. We describe the  plasma using   quantities defined in the co-expanding frame, in which the lepton momenta distribution is isotropic and which is initially related to the merger frame by a Lorentz factor given by Eq.~\ref{eq: initial boost}. In this frame, the initial temperature is directly related to the dark photon mass, $T \approx m'/6$, since in that frame each electron and positron generated by the dark photon has approximately $m'/2$ energy, with a small spread related to the dark photon boost distribution. The initial number density is given by
\begin{equation}
    n_{\text{e}}\approx\frac{N_{\text{tot}}\gamma_0}{4\pi d^3} \, ,
\end{equation}
where we take
\begin{equation}
    N_{\text{tot}} = \frac{dN}{dVdt}\times V_{\text{emit}}t_{\text{emit}}\left(e^{-1000 \text{km}/d}-e^{-1}\right)
\end{equation}
 as the number of dark photons which decay between 1000 km and one decay length from the merger remnant.  $V_{\text{emit}}$ and $t_{\text{emit}}$ are the volume and time over which dark photons are emitted, respectively.  As in the standard fireball model~\cite{Piran:1999kx,Meszaros:2006rc}, when the plasma energy is dominated by relativistic particles, the density evolves as $\rho \propto r^{-4}$, where $r$ is the distance of the plasma shell from the remnant (in the remnant's rest frame), the Lorentz factor of the shell increases as $\gamma \propto r$, and the width in the merger frame, which also determines the duration of the signal, remains constant. If expansion is the only processes which changes the total number density  then the total number of particles is conserved, the total number density scales as $n \propto r^{-3}$, and the temperature goes as $T\propto r^{-1}$.

The electron/positron densities in the plasma can be large enough for pair annihilation, $e^+ e^- \rightarrow \gamma \gamma$, to be very efficient. In this case, the pair creation and annihilation processes quickly leads the number densities of electrons and photons to be related by detailed balance,
\begin{equation}
\label{eq:detailed balance}
    \frac{n_e}{n_\gamma} = \frac{n_e^\text{eq}}{n_\gamma^\text{eq}} \, .
\end{equation}
The RHS of Eq.\ref{eq:detailed balance} refers to the equilibrium number densities for leptons and photons. In this regime, the photon mean free path will be short and the plasma is optically thick. An observable signal emerges once the lepton number density becomes low enough for the plasma to become optically thin, allowing photons to escape. 
Photon bremsstrahlung from electron and positron scattering can also play an important role in the dynamics of the plasma. While pair creation/annihilation preserves the total number of particles, bremsstrahlung increases the total number of particles and consequently the total number density.  Energy conservation implies that the temperature decreases as the number density goes up, thus, bremsstrahlung can lower the peak energy of the photon signal.

To determine if pair annihilation and bremsstrahlung are important effects, we compare their rates in the plasma frame to the number density dilution rate coming from the expansion, 
\begin{equation}
    \Gamma_\text{exp} = -3 \gamma /r \, .
    \label{eq:expansion-rate}
\end{equation}
The rate for pair annihilation is given by~\cite{1982ApJ...258..321S}
\begin{equation}
    \Gamma_\text{annih} \approx \frac{\pi n_e \alpha^2}{m_e^2} \left(1+\frac{2 (T/m_e)^2}{1+\log\left(\frac{2 T}{m_e e^{\gamma_E}}+1.3\right)}\right)^{-1} \! ,
\end{equation}
where $\gamma_E$ is the Euler-Mascheroni constant. The photon production rate from $e^+ e^-$ (or $e^-e^-$) bremsstrahlung in a relativistic gas ($T\gg m_e^2)$ is~\cite{ALEXANIAN:1968zz}
\begin{equation}
\begin{aligned}
    \Gamma_\text{brem} & \approx \frac{2 n_e\alpha^3 \log \left( e^{\gamma_E} m_e^2/T^2 \right)}{9 m_e^2} \left[ 12 \log \left( e^{\gamma_E} m_e^2/T^2 \right) \right. \\ 
    & \left. - 84  + 48 \log \left(e^{\gamma_E} m_e/T \right) \right] \, ,
\end{aligned}
\end{equation}
where we imposed an infrared cutoff on the photon energy $\omega_\gamma > m_e^2/T$, corresponding to photons that can exchange $\mathcal{O}(T)$ of energy in a single scattering and thus quickly thermalize.  For $T \lesssim 1 \, \Mev$, we switch to the non-relativistic rate (for the dominant $e^+ e^-$ case) \cite{1985A&A...148..386H}  
\begin{equation}
    \Gamma_{\text{brem}} \approx \frac{64}{3\sqrt{\pi}}\frac{n_e\alpha^3}{ \sqrt{Tm_e^3}} \, ,
\end{equation}
where we imposed an infrared cutoff $\omega_\gamma>T$, again corresponding to photons that can exchange $\mathcal{O}(T)$ energy in a single scattering. If the annihilation and bremsstrahlung rates are both greater than the expansion rate at the beginning of the plasma evolution, right after most dark photons have decayed, then the plasma thermalizes quickly. Since $n_\gamma \ll T^3$, these processes increase the total number density and decrease the temperature before any significant expansion until $T\leq m_e$, at which point both annihilation and bremsstrahlung rates decay exponentially due to the loss of leptons. Dark photons with masses above $\sim 10$ MeV, have another mechanism for thermalizing rapidly.  Even if the pair annihilation rate is initially slower than the expansion, the energy loss rate from bremsstrahlung \cite{ALEXANIAN:1968zz}
\begin{equation}
    \frac{d \log \rho_e}{dt}\Big{|}_\text{brem} =- \frac{8 n_e\alpha^3 }{m_e^2}\left(\text{Log}\left(\frac{2 T}{e^{\gamma_E}m_e}\right)+\frac{5}{4}\right)
\end{equation}
can be faster than that due to expansion
\begin{equation}
      \frac{d \log \rho_e}{dt}\Big{|}_\text{exp} =- \frac{4\gamma }{r} \, ,
\end{equation}
where $ \rho_e=3 T n_e$ is the energy density of the shell in the shell frame immediately after the dark photons decay. This energy loss decreases the plasma temperature (with insignificant change to the lepton number density), increasing the pair annihilation rate sufficiently to make it faster than the expansion. Afterwards, the fireball evolves as if both pair annihilation and bremsstrahlung were already efficient when the plasma first formed. 

The resulting photon spectrum in the observer frame is approximately thermal, with an apparent temperature given by $\gamma_* T_*$, where $\gamma_*$ and $T_*$ are respectively the plasma's Lorentz factor and temperature when it becomes optically thin for photons. The photon interaction rate drops exponentially once the plasma temperature is below $m_e$ because the electron density becomes Boltzmann suppressed, leading to $T_* \sim m_e/10$ with only a mild logarithmic sensitivity to the initial conditions of the plasma shell. In the thermalized scenario, the temperature drops before the plasma expands significantly, and thus $\gamma_* = \gamma_0 \lesssim 10$, and effectively all of the energy radiated in dark photons gets converted into photons with energies in the $10-1000$ keV range. The photon signal duration is set by light crossing time of the plasma shell, about 0.1-100 s depending on the dark photon parameters. The luminosity of the signal can be estimated by dividing the total energy output by the width of the plasma shell as shown in Eq. \ref{width}.

\begin{figure*}[t!]
\hspace{-0.5cm}
\includegraphics[width=0.45\textwidth]{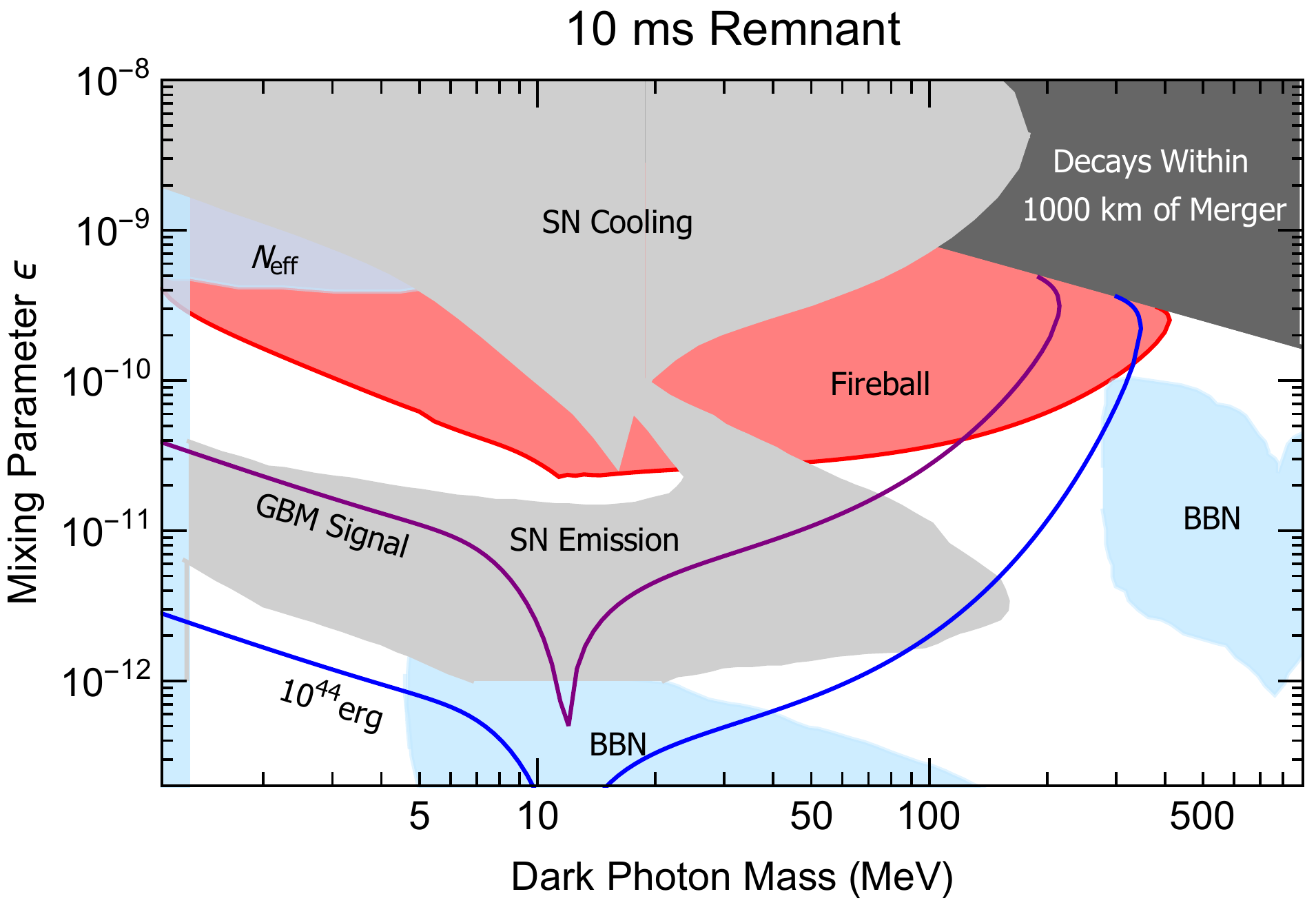}~~~
\includegraphics[width=0.45\textwidth]{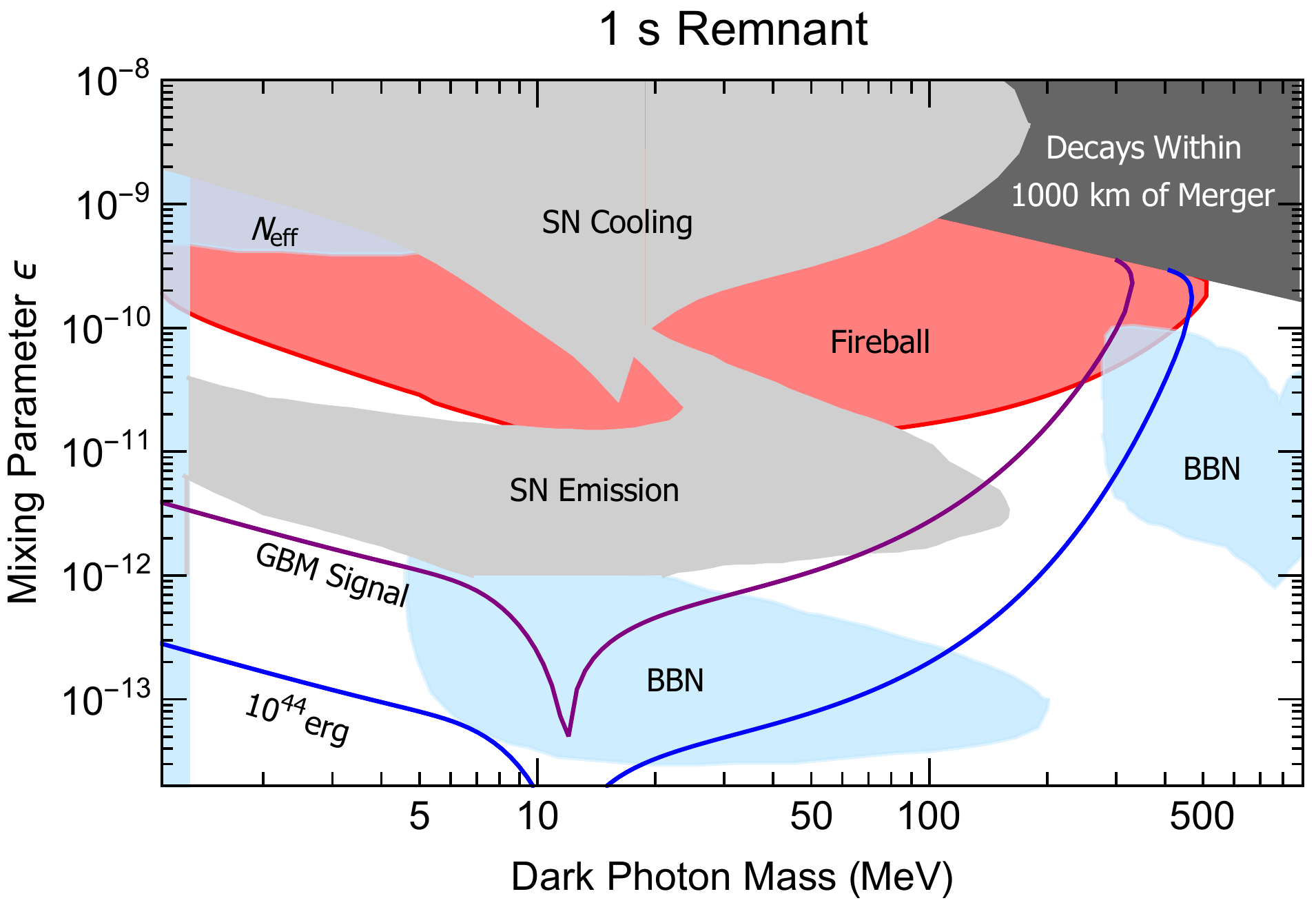}
\caption{The left (right) plot shows the conditions produced when the merger emits dark photons for 10 ms (1 s).  The red region marks where the dark photons form a fireball which produces a thermal spectrum,  below it dark photons generate a dimmer non-thermal signal.  The dark gray region shows where the dark photon decay length is less than 1000 km (close enough to the merger remnant that baryons could affect the signal).  The purple line shows the minimum $\epsilon$ needed to produce a signal visible to Fermi GBM for a merger 100 Mpc away ($\sim 2\times 10^{46}$ ergs), and the blue line shows parameters for a $10^{44}$ erg signal.  The light gray regions marked SN Cooling and SN Emission show the parameter space already ruled out by astrophysical observations \cite{DeRocco:2019njg,Chang:2016ntp}.  The light blue regions labeled $N_{eff}$ and BBN mark space excluded by cosmological observations \cite{Ellis:1990nb, Zhang:2007zzh,Fradette:2014sza,Ibe:2019gpv}. For potential additional complementary constraints see also~\cite{DeRocco:2019njg,Sung:2019xie,Caputo:2022mah}.}
\label{fig:fireball}
\end{figure*}

The parameter space region in which this thermal spectrum occurs is shown in Fig.~\ref{fig:fireball} for two different assumptions about the duration of the remnant. It also shows two curves  which mark the total energy emitted in dark photons.  One curve shows where the signal would be detectable at the Fermi gamma-ray burst monitor (GBM), assuming all of the energy gets converted to photons in the sensitivity range of the instrument ($100-2000$ keV) and a merger distance of $100$ Mpc. The other curve shows the region where the total energy emitted in dark photons is $10^{44}$ ergs as a potential target for future detectors. If the ``fireball'' forms, the dark photon energy would get converted to the GBM range because the initial boost $\gamma_0$ is $\mathcal{O}(1)$ for all relevant masses. This shows that one could probe most of the remaining parameter space above $\epsilon \approx 10^{-11}$, motivated by freeze-in dark matter scenarios, for $m' \approx 1-100$ MeV by searching for gamma-ray signals which coincide with a neutron star merger.

\section{Discussion and future prospects}
\label{sec:Discussion}

The LIGO/Virgo collaboration is currently sensitive to neutron star mergers within $\sim 100$ Mpc of Earth~\cite{2020arXiv200801301I}. The first BNS merger detection, GW170817, was about half this distance,  unusually close given the estimated rate for BNS mergers. There was an associated gamma-ray signal observed shortly after by Fermi-GBM and Integral~\cite{Goldstein:2017mmi}, which has largely confirmed the expectation that BNS mergers are responsible for short gamma-ray bursts (sGRB). Making a discovery of dark photons using the gamma-ray signal discussed in this paper would require distinguishing the dark photon signal from usual sGRBs. This can, in principle, be done using the fact that sGRBs are expected to be highly beamed since they are produced by a relativistic jet, leading to large variations of the observed luminosity from mergers at similar distances. However, potential off-axis gamma-ray emission is still not fully understood, and might be an important background for our proposal~\cite{Lazzati:2016yxl,Gottlieb:2017pju,Lazzati:2020avb}. Even if there is a significant background from off-axis emission, one can use information about the spectrum, signal arrival time and duration as important features to distinguish our signal from sGRBs (in the Supplementary Materials~\cite{supplement} we show how timing can be used to put constraints on the dark photon model). 

The LIGO/Virgo collaboration will be sensitive to neutron star mergers as far as $\sim200$ Mpc when it reaches design sensitivity in the near future~\cite{TheLIGOScientific:2014jea}, at which point it will detect multiple mergers per year. Having large merger statistics, and utilizing realistic remnant profiles from simulations, would allow one to probe the parameter space in Fig.~\ref{fig:fireball} above the purple line and within the fireball region assuming better understanding of the backgrounds from sGRBs. This would cover a large portion of the remaining parameter space motivated by dark matter freeze-in, $\epsilon \gtrsim 10^{-11}$~\cite{Chu:2011be}.

There are two challenges to probing the parameter space in which a fireball never forms. One is modeling the multiple processes that produce photons from a dilute electron positron plasma. We leave the required detailed analysis of this non-thermal emission to future work, but we expect that for part of the parameter space, the main signal will be in $\sim 10$ MeV photons, for which we currently do not have very good coverage. There are a number of proposals such as e-ASTROGRAM \cite{DeAngelis:2016slk}, AMEGO \cite{McEnery:2019tcm}, and MeVCube \cite{10.1117/12.2561510} that target the low MeV range which would significantly increase our reach to dark photons from BNS. The other challenge is that, for smaller couplings the lepton density after decay is smaller and only a reduced fraction of the energy gets converted to photons, making the signal dimmer. One promising direction to compensate the low photon luminosity is to use detectors with better angular resolution, which decreases the background to the signal. Future proposed GW mid-band detectors such as AMIGO \cite{Ni:2019nau}, MAGIS \cite{2021arXiv210402835A}, AION \cite{Badurina:2019hst}, ELGAR \cite{Canuel:2019abg} should be able to detect BNS merger events in advance and with better localization, allowing future  x-ray and gamma-ray telescopes with narrower fields of view to observe the merger.

We have shown that neutron star mergers can be used to search for unstable dark sector particles. We demonstrated this by studying the gamma-ray signatures that arise from dark photon decays, and showed that this can probe a large portion of unconstrained parameter space, including much of the remaining parameter space motivated by dark matter freeze-in. Utilizing the full potential of BNS as probes of dark sectors requires a better understanding of the remnant dynamics and further investigation of how to distinguish the signal arising from dark sectors from the that expected from sGRBs associated with BNS mergers.

\begin{acknowledgments}
The authors thank Tim Dietrich, Julian Krolik, and David Radice for useful discussions and Daniel Ega\~na-Ugrinovic for comments on the draft. GMT was supported in part by the NSF grants PHY-1914480, PHY-1914731, by the Maryland Center for Fundamental Physics (MCFP) and by the US-Israeli BSF Grant 2018236.  MD was supported in part by NSF grant PHY-1818899.
\end{acknowledgments}

\bibliography{ref-draft1}

\clearpage
\newpage
\maketitle
\onecolumngrid
\begin{center}
	\textbf{\large Visible dark photon flashes from neutron star mergers} \\ 
	\vspace{0.05in}
	{ \it \large Supplementary Material}\\ 
	\vspace{0.05in}
	{}
	{Melissa Diamond and Gustavo Marques-Tavares}
	
\end{center}

\setcounter{equation}{0}
\setcounter{figure}{0}
\setcounter{table}{0}
\setcounter{section}{0}
\renewcommand{\theequation}{S\arabic{equation}}
\renewcommand{\thefigure}{S\arabic{figure}}
\renewcommand{\thetable}{S\arabic{table}}
\newcommand\ptwiddle[1]{\mathord{\mathop{#1}\limits^{\scriptscriptstyle(\sim)}}}

\renewcommand{\theHequation}{Supplement.\theequation}
\renewcommand{\theHtable}{Supplement.\thetable}
\renewcommand{\theHfigure}{Supplement.\thefigure}

\section{Dark photon production calculation}

The differential number flux of dark photons is given by:
\begin{equation}
\frac{dN}{dVdt} = \int d\omega\frac{dN}{dVdtd\omega} = \int \frac{d\omega \omega^2 v}{2\pi^2}e^{-\omega/T}(\Gamma_{\text{abs},L}^{'} +2\Gamma_{\text{abs},T}^{'})
\end{equation}
Where $\omega$ is the frequency of the dark photon, $v$ is its velocity and  $\Gamma_{\text{abs},T/L}^{'}$ is the absorptive width of the transverse/longitudinal mode.  Inverse proton-neutron bremsstrahlung is the dominant absorption process in the remnant, so that will be the one considered here. Using the soft radiation approximation the absorptive width is given by~\cite{Rrapaj:2015wgs,Chang:2016ntp}
\begin{equation}
\Gamma^{'}_{\text{ibr},L|T} = \frac{32}{3\pi}\frac{\alpha(\epsilon_m)^2n_{\text{n}} n_{\text{p}}}{\omega^3}\left(\frac{\pi T}{m_{\text{N}}}\right)^{3/2}\langle\sigma_{\text{np}}^{(2)}(T)\rangle \left(\frac{m'^2}{\omega^2}\right)_L \, ,
\end{equation}
where $n_{\text{n}}$ is the number density of neutrons, $n_{\text{p}}$ is the number density of protons, $m_{\text{N}}$ is the neutron mass, $m'$ is the dark photon mass, $T$ is the background temperature, $\langle\sigma_{\text{np}}^{(2)}(T)\rangle$ is the thermal-averaged proton neutron dipole scattering cross section, $(\epsilon_m)^2$ is the in medium mixing angle and $\left(\frac{m'^2}{\omega^2}\right)_L$ only applies to the longitudinal term and is $1$ for the transverse polarizations. The cross section $\langle\sigma_{np}^{(2)}(T)\rangle$ is taken from Ref.~\cite{Rrapaj:2015wgs}.

Plasma effects change the mixing between the photon and the dark photon.  This can be taken into account by using an effective mixing parameter~\cite{Chang:2016ntp}
\begin{equation}
(\epsilon_{m})^{2}_{L|T} = \frac{\epsilon^{2}}{\left(1-\text{Re}\Pi_{L|T}/m'^{2}\right)^{2}+\left(\text{Im}\Pi_{L|T}/m'^{2}\right)^{2}} \, ,
\end{equation}
where $\Pi$ is the polarization tensor.  Its real component is given by:
\begin{equation}
\text{Re}\Pi_{L} = \frac{3\omega_p^2}{v^2}\left(1-v^2\right)\left[\frac{1}{2v}ln\left(\frac{1+v}{1-v}\right)-1\right] 
\end{equation}
\begin{equation}
\text{Re}\Pi_{T} = \frac{3\omega_p^2}{2v^2}\left[1-\frac{1-v^2}{2v}ln\left(\frac{1+v}{1-v}\right)\right] \, .
\end{equation}
Here the velocity of the dark photon is determined by $v = \sqrt{1-\frac{m'^2}{\omega^2}}$, and $\omega_p$ is the plasma frequency, which for the remnant is well approximated by the one for a gas of degenerate electrons given by
\begin{equation}
\omega_p^2 = \frac{4\pi \alpha n_e}{\sqrt{m_e^2+\left(3\pi^2n_e\right)^{2/3}}} \, ,
\end{equation}
where $m_e$ and $n_e$ denote the electron mass and number density respectively.  

Within the merger, standard model photons are in local thermal equilibrium.  Thus, the imaginary part of the polarization becomes:
\begin{equation}
\text{Im}\Pi_{L|T} = -\omega\left(1-e^{-\omega/T}\right)\Gamma_{abs,L|T} \, ,
\end{equation}
where $\Gamma_{abs,L|T}$ is the absorptive width of the standard model photon, taken to be
\begin{equation}
\Gamma_{ibr,L|T} = \frac{32\alpha}{3\pi}\frac{n_n n_p}{\omega^3}\left(\frac{\pi T}{m_N}\right)^{3/2}\langle\sigma_{np}^{(2)}(T)\rangle\left(\frac{m'^2}{\omega^2}\right)_L \, ,
\end{equation}
\begin{equation}
\Gamma'_{abs} = (\epsilon_m)^2_{L|T}\Gamma_{abs} \, .
\end{equation}

\section{Gravitational Redshift} \label{App: Redshifting}
The number of dark photons emitted, their Lorentz factor upon decaying and the total energy stored in the expanding fireball shell will be altered by redshifting effects as the dark photons climb out of the gravitational well created by the merger remnant. To compute the exact impact of gravitational redshift on the signal requires a more accurate picture for the density profile of the remnant than the one we used to compute the signal.  Redshifting will reduce the number of dark photons that escape the merger, and will reduce the total energy of the photos that do.  Still assuming spherical symmetry, the number of dark photons that escape would now be found by taking
\begin{equation}
\frac{dN}{dt} = 4\pi\int_{r_{min}}^{r_{max}}r^2 dr\int_{\omega_m}^{\infty} \frac{d\omega \omega^2 v}{2\pi^2}e^{-\omega/T}(\Gamma_{\text{abs},L} +2\Gamma_{\text{abs},T}),
\end{equation}
Where $\omega_m$ is the minimal energy a dark photon needs to escape the merger
\begin{equation}
\omega_m = \frac{m'}{\sqrt{1-2\left(\frac{GM_{int}(r)}{r}\right)}} \, ,
\end{equation}
where $M_{int}(r)$ is the total mass contained within the radius $r$ of the merger remnant. The total energy carried away by the dark photons can be found using
\begin{equation}
\frac{dE}{dt} = 4\pi\int_{r_{min}}^{r_{max}}r^2 dr\int_{\omega_m}^{\infty} \frac{d\omega \omega^3\sqrt{1-2\left(\frac{GM_{int}(r)}{r}\right)} v}{2\pi^2}e^{-\omega/T}(\Gamma_{\text{abs},L} +2\Gamma_{\text{abs},T}) \, .
\end{equation}

\begin{figure}[t]
	\centering
	\includegraphics[scale=0.5]{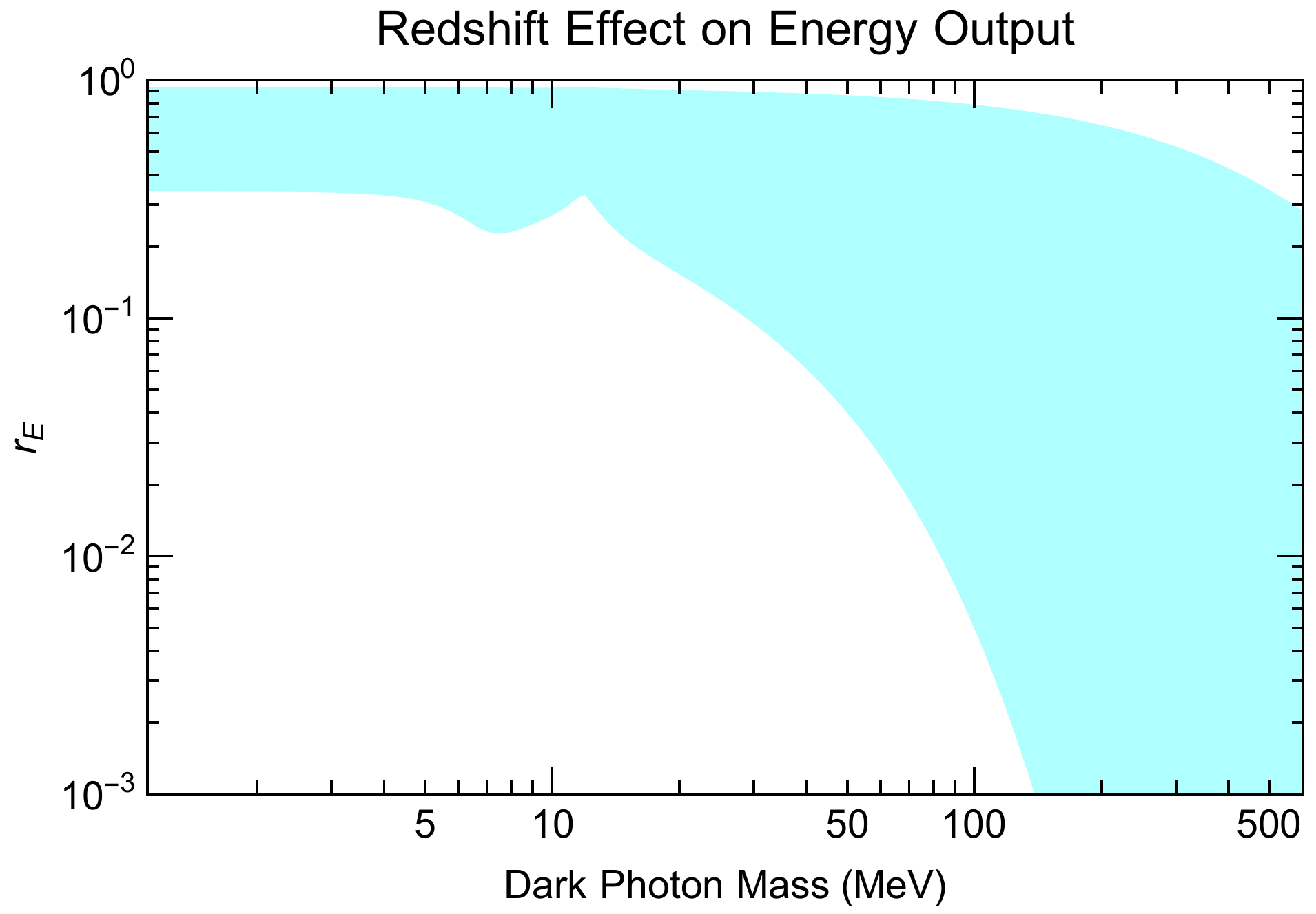}
	\caption{Range of possible ratios between dark photon luminosity when redshifting effects are included and when such effects are neglected for a $2.7M_{\odot}$ merger remnant.  The upper limit for this ratio comes from a model which assumes the remnant has a constant density throughout of $4\times 10^{14}$ g cm$^3$.  The lower limit on this ratio comes from a model where the remnant has density of $4\times 10^{14}$ g cm$^3$ in the dark photon emission region (5 km-10 km from the center), with the remaining mass of the remnant stored in the inner 5 km.  This strongly suppresses the dark photon luminosity because the amount of mass in the innermost region would be enough to form a black hole, preventing many dark photons from escaping.}
	\label{fig:redshifting}
\end{figure}

The average energy of the dark photons, $E_{ave}$, is found by dividing the total energy output after accounting for redshifting by the total number of dark photons that successfully escape the merger.  In Fig.~\ref{fig:redshifting} we show the approximate range of the ratio between the dark photon luminosity  when redshifting effects are considered and when such effects are neglected for a $2.7M_{\odot}$ merger remnant, $r_E$.   The upper bound of this ratio is for a model that assumes the remnant has a constant density of $4\times10^{14}$ g cm$^{-3}$ throughout, extended out to a radius such that the total mass is $2.7M_{\odot}$.  The lower bound is set by a model where the remnant has a density of $4\times10^{14}$ g cm$^{-3}$ in the emission region (between 5 km and 10 km from the merger center), no density in the region beyond 10 km from the core, and the remaining mass of the 2.7 $M_{\odot}$ remnant residing in the inner 5 km of the remnant.  Most models of merger remnants have similar densities to those used here in the emission region, and become more dense closer to the core~\cite{Perego:2019adq,Radice:2020ids,Bernuzzi:2020txg,Endrizzi:2019trv}, however the density in the lower curve is so large that it would already be a black-hole which is not the case for the simulations. The first model underestimates the central region's density and thus leads to a shallower gravitational potential.  The second model, on the other hand, significantly overestimates the density in the central region, leading to a large escape velocity and thus a significant decrease in luminosity due to gravitational redshifting. The figure shows how sensitive to the density modeling this effect is, specially when considering dark photon masses larger than the remnants temperature. The real effect, using a more realistic density profile, will be somewhere in between the two limiting cases, but would primarily affect the large mass end of the plot.

\section{Ejecta from Binary Neutron Star Merger}

\begin{figure}[t]
	\centering
	\includegraphics[scale=0.6
	]{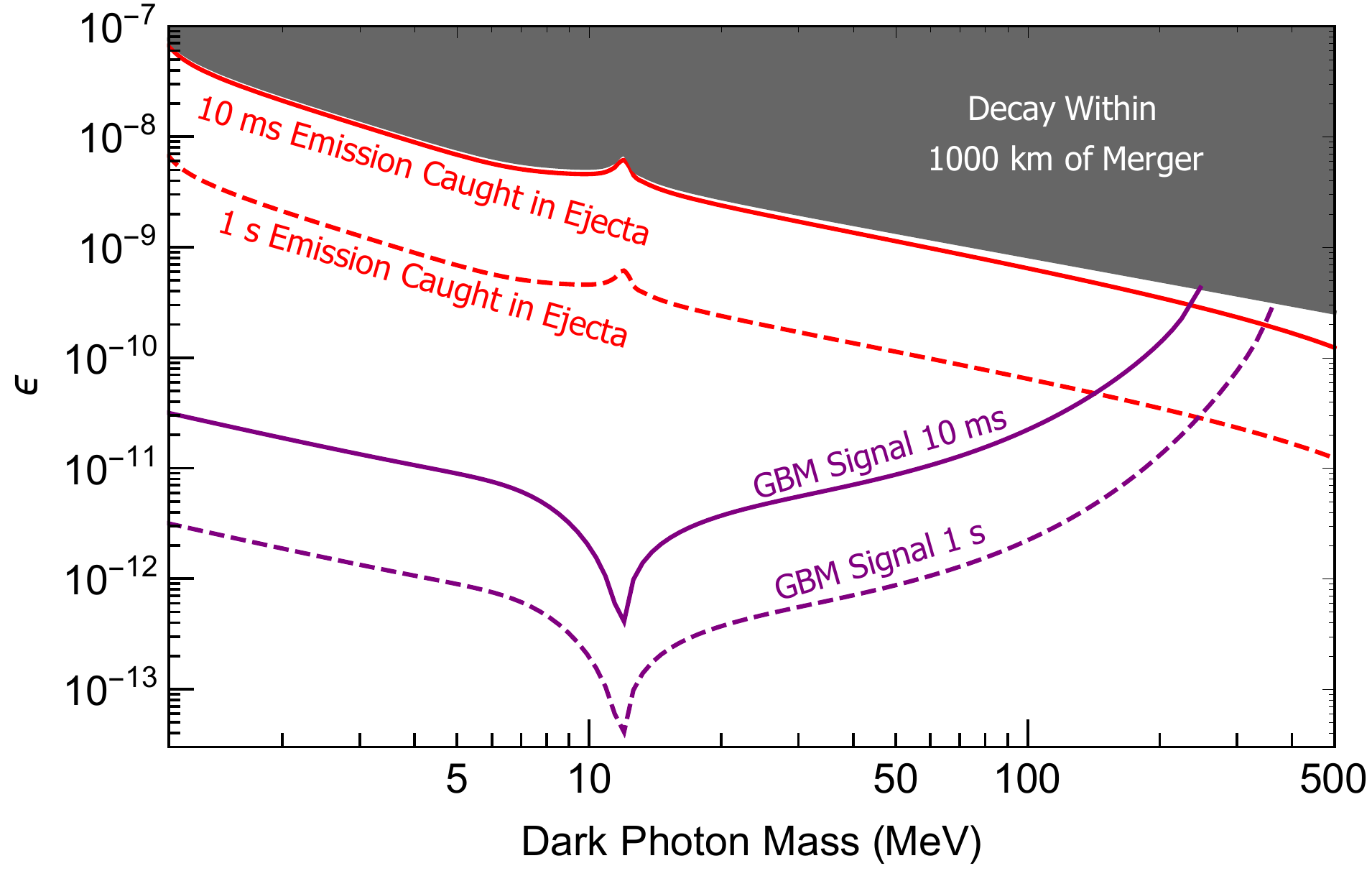}
	\caption{The above plot shows the parts of dark photon parameter space which may be affected by fast moving ejecta.  The red lines (solid for 10 ms, dashed for 1 s) mark the minimum $\epsilon$ where fast moving ejecta begin to affect the observed gamma ray signal.  For reference the original region where dark photons decay within 1000 km of the merger remnant and produce no distinct signal is marked in dark gray and the minimum couplings needed for the produced gamma-ray signal to be visible by Fermi GBM from 100 Mpc away are shown in purple (solid for 10 ms and dashed for 1 s).  Ejecta will only impact signals from dark photons with the highest masses. }
	\label{fig:ejecta}
\end{figure}

In the moments after a Binary Neutron Star Merger, a small fraction of the remnant mass can be ejected with large velocities of around 0.1-0.3~\cite{Hajela:2019mjy, Radice:2018pdn}. Even though dark photons can pass through this ejecta without interacting, it may affect the fireball dynamics and photon signal if dark photons decay before they have passed through the ejecta.  The average dark photon produced in the merger moves faster than 0.3 for all of the relevant parameter space, and so some fraction of the dark photons will always outrun the ejecta. Depending on how long the dark photon emission lasts, dark photons emitted at late times may not be able to outrun the ejecta by the time they decay, and our approximation of ignoring baryon matter in the fireball would not be realistic.

In order to investigate the regions of parameter space that could be affected by the presence of the fast ejecta, we found the parameters for which a dark photon with average velocity would outrun the fast ejecta before decaying considering the case when the dark photon is emitted 10 ms and 1 s after the merger (which correspond to the last dark photons emitted in the two scenarios considered in the paper). The results are shown in Fig.~\ref{fig:ejecta}. Even for remnants that can emit dark photons for a total period of $\Delta t$, we can break up the emission in smaller time intervals $\delta t$, and because the average dark photon is faster than the ejecta, there will always be some small interval for which the average dark photon will outrun the ejecta and our analysis ignoring baryons holds (due to the spread in velocity of the dark photon, shells emitted at different times can reach one another, but for the purposes of our analysis, it only matters whether most of the particles in that shell outrun the ejecta or not). One can see from Fig.~\ref{fig:ejecta} that for the 10 ms emission case, the requirement that the decay occurs at least 1000 km away from the remnant is almost enough to guarantee that most dark photons will outrun the ejecta, since the small region in between the two curves corresponds to the largest couplings we are sensitive to, considering a smaller emission time to guarantee that all dark photons outrun the baryons would lead to a significant change in sensitivity. For the 1 s emission scenario, note that we only need to worry about the region of parameter space which would be covered by a 1 s emission but not for a 10 ms emission, since from our previous argument we could focus on the signal coming from the first 10 ms of emission of the more long lived remnant. This region corresponds to weaker couplings, which lead to longer decay lengths and thus almost all of it is below the line corresponding to outrunning the ejecta in Fig.~\ref{fig:ejecta}, except for a very small region at corresponding to masses larger than 100 MeV. 
\\
\section{Constraining Power}

There are a variety of properties of the predicted gamma-ray signal produced by dark photons that can differ from the observed gamma-ray signals in binary neutron star mergers.  This can be used to place constraints on the dark photon parameter space. In this section we explore two such properties, the timing of the signal and its duration. 

\begin{figure}
	\centering
	\includegraphics[scale=0.8]{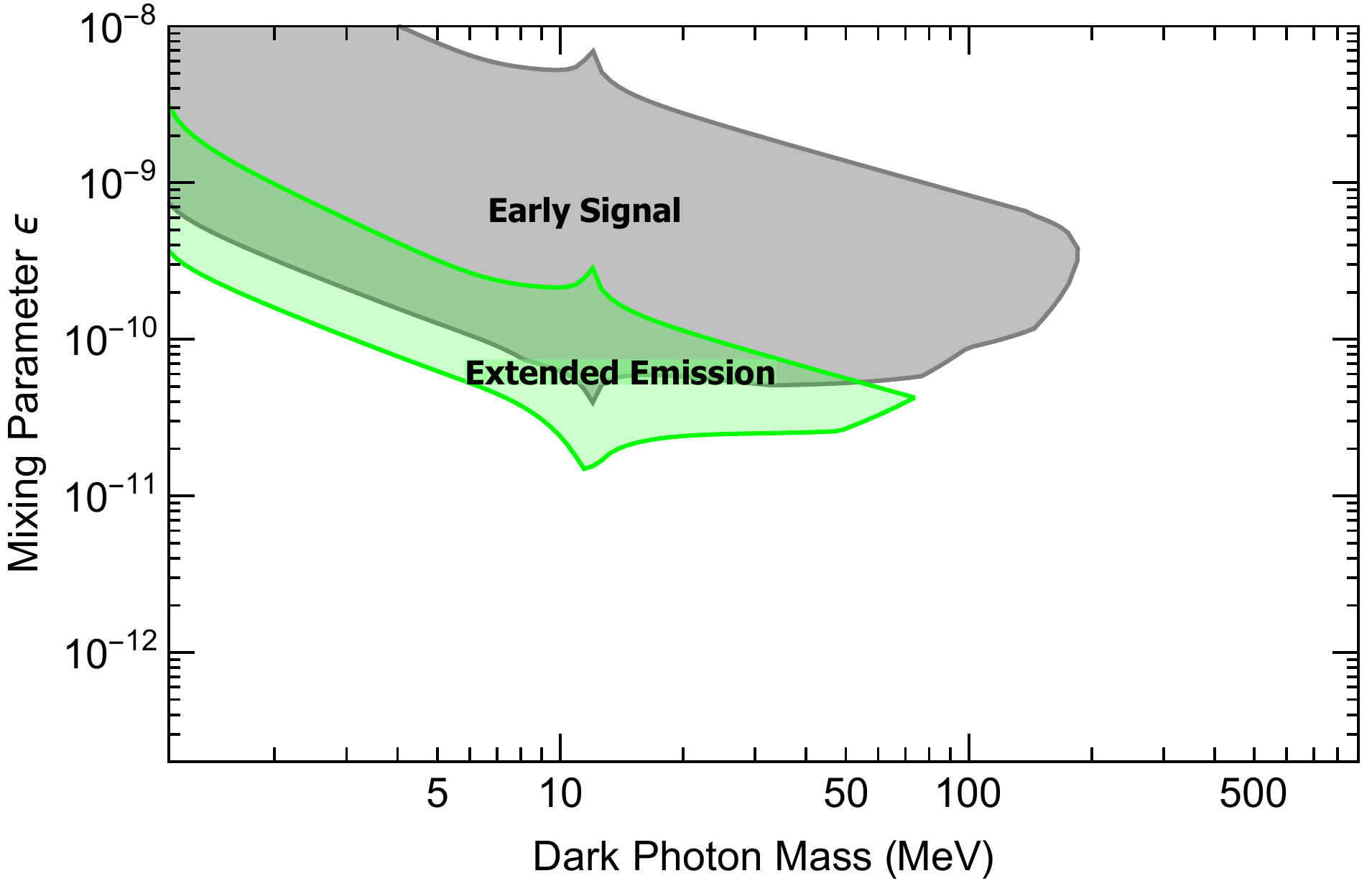}
	\caption{Example parameter space that could be excluded by observed emission from GRB 170817A assuming 10ms emission and the simplified remnant model discussed in the main text.  The gray region is ruled out by dark photons producing excess emission spread out over the time after GRB 170817A.  The green region is ruled out by excess emission which would reach earth between the gravitational wave signal of GW170817 and the gamma-ray signal GRB 170817A.}
	\label{fig:exclusion}
\end{figure}

The gamma-ray burst associated with GW170817, GRB 170817A, is the only GRB observed in association with a BNS merger, and we will use it as a study case to show how one can use specific observations to constrain dark photon parameter space.  As discussed in the main text, dark photons can be produced in the BNM remnant during the time that it is hot, and has not yet collapsed into a black hole.  For the remnant associated with GRB 170817A, this is believed to be at least 10ms \cite{Margalit:2017dij}, and therefore we will use a 10ms emission for our analysis.  The dark photons would stream out from remnant at an average velocity
\begin{equation} \label{eq:velocity}
v=\sqrt{1-\left(\frac{m'}{\langle\omega\rangle}\right)^2} \, ,
\end{equation} 
where $\langle\omega\rangle$ is their average energy.  The much lighter electrons and positrons produced by their decay, would stream at near the speed of light.  The time delay between the gravitational wave signal and the gamma-ray signal, is then the difference between the times it takes the gravitational waves, travelling at the speed of light, and a dark photon, travelling with velocity given by Eq.~\ref{eq:velocity}, to traverse the dark photon decay length,
\begin{equation}
\delta t = \left(\frac{1-v}{v}\right)d \, .
\end{equation}
The gamma-ray burst associated with GW170817 appeared about 1.7 s after the merger~\cite{LIGOScientific:2017ync}.  Additional studies of the burst limited precursor gamma-ray emission to $11\times 10^{-7}$ ergs s$^{-1}$ cm$^{-2}$ for bursts lasting up to 1s and $3.7\times10^{-6}$ ergs s$^{-1}$ cm$^{-2}$ for bursts lasting up to 0.1s in the 200 s leading up to GRB 170817A \cite{Goldstein:2017mmi}.  Using this, we show an example of how one could rule out dark photons which produce a gamma-ray signal detectable to Fermi GBM but which arrives earlier than the observed signal did in Figure \ref{fig:exclusion}.  

A complimentary way to constrain dark photon parameter space is to consider the duration of the observed signal.  The gamma-ray signal from dark photons should persist for the light crossing time of the fireball.  The thickness of the fireball shell, $\delta$, is defined in Eq.(8) of the main text.  Follow-up analysis of GRB 170817A limited the extended emission succeeding the burst averaged over a 10s interval to $6.6\times 10^{-8}$ ergs s$^{-1}$ cm$^{-2}$ for the 100s after the burst \cite{Goldstein:2017mmi}.  We present the regions of dark photon parameter space which would produce excess emission beyond the end of GRB 170817A in  Figure \ref{fig:exclusion}. 

\end{document}